\documentstyle[twoside]{article}

\catcode`\@=11
\long\def\@makefntext#1{
\protect\noindent \hbox to 3.2pt {\hskip-.9pt  
$^{{\eightrm\@thefnmark}}$\hfil}#1\hfill}		

\def\@makefnmark{\hbox to 0pt{$^{\@thefnmark}$\hss}}	
	
\def\ps@myheadings{\let\@mkboth\@gobbletwo
\def\@oddhead{\hbox{}
\rightmark\hfil\eightrm\thepage}   
\def\@oddfoot{}\def\@evenhead{\eightrm\thepage\hfil
\leftmark\hbox{}}\def\@evenfoot{}
\def\sectionmark##1{}\def\subsectionmark##1{}}

\oddsidemargin=\evensidemargin
\newcounter{sectionc}\newcounter{subsectionc}\newcounter{subsubsectionc}
\renewcommand{\section}[1] {\vspace{12pt}\addtocounter{sectionc}{1} 
\setcounter{subsectionc}{0}\setcounter{subsubsectionc}{0}\noindent 
	{\tenbf\thesectionc. #1}\par\vspace{5pt}}
\renewcommand{\subsection}[1] {\vspace{12pt}\addtocounter{subsectionc}{1} 
	\setcounter{subsubsectionc}{0}\noindent 
	{\bf\thesectionc.\thesubsectionc. {\kern1pt \bfit #1}}\par\vspace{5pt}}
\renewcommand{\subsubsection}[1] {\vspace{12pt}\addtocounter{subsubsectionc}{1}
	\noindent{\tenrm\thesectionc.\thesubsectionc.\thesubsubsectionc.
	{\kern1pt \tenit #1}}\par\vspace{5pt}}
\newcommand{\nonumsection}[1] {\vspace{12pt}\noindent{\tenbf #1}
	\par\vspace{5pt}}

\newcommand{\textlineskip}{\baselineskip=13pt}
\newcommand{\smalllineskip}{\baselineskip=10pt}

\def\eightcirc{
\begin{picture}(0,0)
\put(4.4,1.8){\circle{6.5}}
\end{picture}}
\def\eightcopyright{\eightcirc\kern2.7pt\hbox{\eightrm c}} 

\newcommand{\copyrightheading}[1]
	{\vspace*{-2.5cm}\smalllineskip{\flushleft
        {\footnotesize Los Alamos archives: quant-ph/9601019 #1}\\
        {\footnotesize $\eightcopyright$\, H.C. Rosu, Phys. Rev. A 54
        (Oct. 1996) 2571-2576}\\
	 }}

\def\abstracts#1#2#3{{
	\centering{\begin{minipage}{4.5in}\baselineskip=10pt\footnotesize
	\parindent=0pt #1\par 
	\parindent=15pt #2\par
	\parindent=15pt #3
	\end{minipage}}\par}} 

\renewenvironment{thebibliography}[1]
	{\frenchspacing
	 \ninerm\baselineskip=11pt
	 \begin{list}{\arabic{enumi}.}
        {\usecounter{enumi}\setlength{\parsep}{0pt}     
	 \setlength{\leftmargin 12.7pt}{\rightmargin 0pt} 
         \setlength{\itemsep}{0pt} \settowidth
	{\labelwidth}{#1.}\sloppy}}{\end{list}}

\newcounter{itemlistc}
\newcounter{romanlistc}
\newcounter{alphlistc}
\newcounter{arabiclistc}

\def\@citex[#1]#2{\if@filesw\immediate\write\@auxout
	{\string\citation{#2}}\fi
\def\@citea{}\@cite{\@for\@citeb:=#2\do
	{\@citea\def\@citea{,}\@ifundefined
	{b@\@citeb}{{\bf ?}\@warning
	{Citation `\@citeb' on page \thepage \space undefined}}
	{\csname b@\@citeb\endcsname}}}{#1}}

\newif\if@cghi
\def\cite{\@cghitrue\@ifnextchar [{\@tempswatrue
	\@citex}{\@tempswafalse\@citex[]}}
\def\citelow{\@cghifalse\@ifnextchar [{\@tempswatrue
	\@citex}{\@tempswafalse\@citex[]}}
\def\@cite#1#2{{$\null^{#1}$\if@tempswa\typeout
	{IJCGA warning: optional citation argument 
	ignored: `#2'} \fi}}

\def\@refcitex[#1]#2{\if@filesw\immediate\write\@auxout
	{\string\citation{#2}}\fi
\def\@citea{}\@refcite{\@for\@citeb:=#2\do
	{\@citea\def\@citea{, }\@ifundefined
	{b@\@citeb}{{\bf ?}\@warning
	{Citation `\@citeb' on page \thepage \space undefined}}
	\hbox{\csname b@\@citeb\endcsname}}}{#1}}

\def\@refcite#1#2{{#1\if@tempswa\typeout
        {IJCGA warning: optional citation argument
	ignored: `#2'} \fi}}

\def\refcite{\@ifnextchar[{\@tempswatrue
	\@refcitex}{\@tempswafalse\@refcitex[]}}

\def\pmb#1{\setbox0=\hbox{#1}
	\kern-.025em\copy0\kern-\wd0
	\kern.05em\copy0\kern-\wd0
	\kern-.025em\raise.0433em\box0}

\def\fnt#1#2{\footnotetext{\kern-.3em
	{$^{\mbox{\scriptsize #1}}$}{#2}}}

\def\runninghead#1#2{\pagestyle{myheadings}
\markboth{{\protect\footnotesize\it{\quad #1}}\hfill}
{\hfill{\protect\footnotesize\it{#2\quad}}}}
\font\tenrm=cmr10
\font\tenit=cmti10 
\font\tenbf=cmbx10
\font\bfit=cmbxti10 at 10pt
\font\ninerm=cmr9

\font\eightrm=cmr8

\def\qed{\hbox{${\vcenter{\vbox{			
   \hrule height 0.4pt\hbox{\vrule width 0.4pt height 6pt
   \kern5pt\vrule width 0.4pt}\hrule height 0.4pt}}}$}}

\begin{document}

\runninghead{H.C. Rosu $\ldots$} {H.C. Rosu$\ldots$}


\normalsize\textlineskip
\thispagestyle{empty}
\setcounter{page}{1}

\copyrightheading{}			

\vspace*{0.88truein}

\centerline{\bf DARBOUX-WITTEN TECHNIQUES FOR THE DEMKOV-OSTROVSKY PROBLEM}
\vspace*{0.035truein}
\vspace*{0.37truein}
\centerline{\footnotesize HARET C. ROSU}
\vspace*{0.015truein}
\centerline{\footnotesize\it Instituto de F\'{\i}sica,
Universidad de Guanajuato, Apdo Postal E-143, Le\'on, Gto, Mexico}
\centerline{\footnotesize\it rosu@ifug3.ugto.mx}
\baselineskip=10pt
\vspace*{10pt}
\vspace*{0.225truein}

\vspace*{0.21truein}
\abstracts{
The bosonic strictly isospectral problem for Demkov-Ostrovsky (DO) effective
potentials in the radially nodeless sector is first solved in the
supersymmetric Darboux-Witten (DW) half line (or $l$-changing) procedure. As
an application, for the $\kappa =1$ class, if one goes back to optics
examples, it
might be possible to think of a one-parameter family of Maxwell lenses having
the same optical scattering properties in the nodeless radial sector.
Although the
relative changes in the index of refraction that one may introduce in this
way are at the level of several percents, at most, for all DO orbital quantum
numbers
$l\geq 0$, the index profiles are different from the original Maxwell one,
possessing an inflection point within the lens. 
 I pass then to the DW full line (or $N$-changing) procedure, obtaining the
corresponding Morse-type problem for which the supersymmetric results are
well established, and finally come back to the half line with well-defined
results.\\
PACS number(s):  03.65.-w, 11.30.Pb
}{}{}


\textlineskip                  
\vspace*{12pt}                 

\vspace*{1pt}\textlineskip	
\vspace*{-0.5pt}
\noindent



\noindent




\noindent



\section{\bf{INTRODUCTION}} 

Recently, the Schr\"odinger equation at zero energy with the
Demkov-Ostrovsky (hereafter DO) class of focusing
potentials $^{1}$
$$
V_{\kappa}(r)=
-\frac{w{\cal E}_0}{(r/R)^2[(r/R)^{-\kappa}+(r/R)^{\kappa}]^2}
\eqno(1)
$$
has been studied in the supersymmetric quantum-mechanical context $^{2}$.
In Eq.~(1) $w>0$ is a coupling constant, ${\cal E} _0$ is an energy scale,
$R>0$ is the spatial scale of the
potential, $\kappa$ is the Lenz-Demkov-Ostrovsky parameter ($\kappa=1/2$ class
was related
to the atomic aufbau $^{3}$, while the $\kappa=1$ case is a wave approach
to the Maxwell fish-eye lens $^{1}$).
It has been shown that the radially nodeless sector of the
DO wave problem may be worked out in the supersymmetric approach.
As is quite well known, the essence of Witten's supersymmetric
quantum mechanics $^{4}$ (Darboux procedure, in mathematics literature)
is a pair of 
Riccati equations, which, when the particular solution is used, entails the
boson-fermion symmetry at the level of standard one-dimensional
Schr\"odinger equations. Moreover, when use is made of the general Riccati
solution one can solve the isospectral problem, i.e., one is able to
generate, in the most simple procedure, classes of one-parameter families of
potentials having the same eigenvalues as the original potential. The
one-parameter family of {\it fermionic} DO potentials for the zero energy case
has been already obtained
in $^{2}$.

The organization of the paper is as follows.
In Sec. II,  I shall obtain the ``half line"
one-parameter family of
{\it bosonic} DO effective potentials at zero energy, in the nodeless sector,
having the same fermionic superpartner. Moreover, I shall focus on the optical
application, introducing, in the nodeless radial sector, a one-parameter family
of generalized Maxwell indices of
refraction, in the sense that they are defined for each partial wave and depend
on the family parameter $\lambda$. 
Then, in Sec. III, I shall use the Langer transform
to go to the full line Schr\"odinger equation for which the Darboux-Witten
(DW) procedure
applies in a standard way and return to the half line with simple yet
well-defined results. A conclusion section ends up the paper.

\section{{\bf DO-ISOSPECTRAL PROBLEM ON THE HALF LINE}} 

I shall use the
double DW procedure to first delete and then reintroduce a normalizable state
in a spectrum, taking the initial DO Schr\"odinger equation
at zero energy, which for the quantized values $w_{N,\kappa}=(2\kappa)^2
[N+(2\kappa)^{-1}][N+(2\kappa)^{-1}-1]$ is known to have a zero-energy,
regular,
normalizable state whose radial part is ($\rho=r/R$)
$$
R _{Nl}(\rho)=\frac{{\cal N} _{Nl}\rho ^l}
{(1+\rho ^{2\kappa})^{(2l+1)/2\kappa}}
C_{N-1-l/\kappa}^{(2l+1)/2\kappa+1/2}(\xi)~,
\eqno(2)
$$
where $N=n+(\kappa ^{-1}-1)l$, $n=n_r +l+1$, $n_r=0,1,2,...$, are the ``total",
``principal" and ``radial" DO quantum numbers, respectively, $l$ is the
DO orbital one, $C_p^q(\xi(\rho))$ are Gegenbauer polynomials of variable
$\xi=(1-\rho ^{2\kappa})/(1+\rho ^{2\kappa})$, and
${\cal N} _{Nl}$ are the normalization constants that I shall not take into 
account henceforth. The degree of degeneracy
of the zero-bound state is $N^2$ similar to the hydrogen atom.
In the $n_r=0$ sector the
Gegenbauer polynomials are unity for any degree $p$, so they do not enter
the following.

To generate the bosonic
isospectral family I need the general solution of the fermionic Riccati
equation
$$
U^+_{eff}=+\frac{dW_{\kappa}(\rho)}{d\rho}+W^2_{\kappa}(\rho)~,
\eqno(3)
$$
for which the particular solution is the following $^{2}$
$$
W_{\kappa}(\rho)=\frac{l}{\rho}-\frac{2l+1}{\rho (1+\rho ^{2\kappa})}~,
\eqno(4)
$$
and where $U^+_{eff}$ is the fermionic effective superpartner $^{2}$.
Substitution of the general solution supposed to be of the form
${\cal W}_{\kappa} ={\cal V} ^{-1}_{\kappa} +W_{\kappa}$ in Eq.~(3) leads to
the following simple differential equation for ${\cal V}_{\kappa} $
$$
-\frac{d{\cal V}_{\kappa}}{d\rho}+2W_{\kappa}{\cal V}_{\kappa}=-1
\eqno(5)
$$
with the general solution (the particular superpotential is the negative
logarithmic derivative of the radial factor $f_{\kappa}$)
$$
{\cal V} _{\kappa ,\lambda}= f^{-2}_{\kappa}\left(\lambda+
\int _{0}^{\rho} f^2_{\kappa}(\rho ^{'})d\rho ^{'}\right)~,
\eqno(6)
$$
where $\lambda$ is a positive, real parameter, and
the DO radial factor in the $n_r=0$ sector is $^{2}$
$$
f_{\kappa}(\rho)=\frac{\rho ^{l+1}}{(1+\rho ^{2\kappa})^{(2l+1)/2\kappa}}~.
\eqno(7)
$$
Denoting the integral in Eq.~(6) by $I_{0;\kappa}(\rho)$ the general
superpotential reads
$$
{\cal W} _{\lambda,\kappa}=W_{\kappa}+\frac{d}{d\rho}
\ln[I_{0;\kappa}(\rho)+\lambda]~,
\eqno(8)
$$
as one can show easily. Then, the isospectral effective potential
family can be written in terms of the original effective DO potential and
the radial factor $f_{\kappa}$ as follows
$$
U_{bos,\kappa}(\rho;\lambda)=U^{-}_{\kappa}(\rho)-
2[\ln(I_{0;\kappa}+\lambda)]^{''} =U^{-}_{\kappa}(\rho)-
\frac{4f_{\kappa}f_{\kappa}^{'}}{I_{0;\kappa}+\lambda}+
\frac{2f_{\kappa}^4}{(I_{0,\kappa}+\lambda)^2}~,
\eqno(9)
$$
the primes denoting derivatives with respect to $\rho$.
It remains to calculate the integral $I_{0;\kappa}$ in order to consider
the problem as solved. For that, the trigonometric representation is the
most convenient. With $\rho =[\tan\beta]^{1/\kappa}$ the integral turns into
$$
I_{0;\kappa}=\frac{1}{\kappa}\int_0^{\beta}[\cos\beta]^{-\frac{2}{\kappa}}
[\sin \beta]^{\frac{2l+3-\kappa}{\kappa}}d\beta~.
\eqno(10)
$$
Then, for the physical cases, 
$\kappa=1/2$ and
$\kappa=1$, one can use the formulas 2.524.1
and 2.518.1, respectively, in $^{5}$, as the most compact ones,
in order to get
$$
I_{0;\frac{1}{2}}=2\int_0^{\beta}
\frac{[sin\beta]^{4l+5}}{[\cos\beta]^4}d\beta=2
\sum_{k=0}^{2l+2}(-1)^{k+1}\frac{(2l+2)!}{k!(2l+2-k)!}
\frac{[\cos\beta]^{2k-3}}{[2k-3]}
\eqno(11)
$$
and
$$
I_{0;1}=\int_0^{\beta}\frac{[sin\beta]^{2l+2}}{[\cos\beta]^2}d\beta=
\frac{[\sin\beta]^{2l+1}}{\cos\beta}-(2l+1)\int_0^{\beta}
[\sin\beta]^{2l}d\beta~.
\eqno(12)
$$
Finally, for the last integral in Eq.~(12) I used the formula 2.513.1
in $^{5}$ leading to
$$
I_{0;1}=  
\frac{[\sin\beta]^{2l+1}}{\cos\beta}-
\frac{(2l+1)}{4^l}\frac{(2l!)}{(l!)^2}\beta  -2(2l+1)
\frac{(-1)^{l}}{4^l}\sum_{k=0}^{l-1}(-1)^{k}\frac{(2l)!}{k!(2l-k)!}
\frac{[\sin(2l-2k)\beta]}{[2l-2k]}~.
\eqno(13)
$$

\subsection{{\bf Application}} 

The interesting physical point may be seen more directly in the Maxwell
fish-eye case, $\kappa =1$. Notice that now $N=n$.
Suppose we use an optical physics example.
Then, one might think
of a whole family of indices of refraction of the Maxwell type, i.e., a
family of lenses with the same remarkable optical properties.
Since after performing the double DW procedure the centrifugal
barrier is the same for the whole bosonic family and equal to that of
$U^{-}_{eff}$, $^{6}$ following Demkov and Ostrovsky $^{1}$,
to the bosonic family of Maxwell fish-eye potentials
$$
V_{1}(\rho; l,\lambda)= -\frac{(2l+1)(2l+3)}{(1+\rho ^2)^2}-
\frac{4f_{1}f_{1}^{'}}{I_{0;1}+\lambda}+
\frac{2f_{1}^4}{(I_{0;1}+\lambda)^2}
\eqno(14)
$$
would correspond the family of Maxwell indices
$$
n_1(\rho; l,\lambda)\propto \sqrt{-V_{1}(\rho; l,\lambda)}~.
\eqno(15)
$$
The one-parameter family of Maxwell lenses that might be thought of in this way
are identical in the sense that, quantum-mechanically, the scattering matrices
are identical $^{6}$. I emphasize that all these $n(\rho; l,\lambda)$
are valid in the $n=l+1$ sector.(See Figs. 1 and 2.)

It is known in the literature $^{6}$ that pair combinations (including
double-ones, as the double DW used here) of all the
three known procedures to delete and reintroduce a nodeless, normalizable
state in a spectrum, i.e., the Darboux-Witten, the Abraham-Moses, and the
Pursey ones,
generate five distinct, one-parameter, ``isospectral" families of (effective)
potentials. However, the supplementary four are not {\em exactly} isospectral.
Nevertheless, their
quantum-mechanical scattering matrices are related by simple relationships
$^{6}$.

Since I have used the double DW procedure I first deleted and then reinserted 
the zero-energy
bound state, or a sector of it. The price for such an action is the damping of
the new radial factors as follows
$$
f_{bos, \kappa}(\rho)=\frac{f_{\kappa}(\rho)}{I_{0,\kappa}+\lambda}~.
\eqno(16)
$$
From the plots I did, one can notice that the ``physical" effect of the
family parameter $\lambda$ on the index of refraction, though not
important - it is at the level of several percents 
and goes down strongly with increasing $l$ -
is sufficient to generate an inflection point in the index profile. Its
physical nature should be further examined.
On the other hand, $\lambda$ is acting as a huge damping parameter
for the radial factor.

\section{{\bf FROM THE HALF LINE TO THE FULL LINE AND BACK}} 

What I have discussed so far has to do with the half-line (radial) problem.
It is a supersymmetric construction which for the very close hydrogen problem
has been questioned by Haymaker and Rau $^{7}$. This is so because the
interpretation of the DW construction along the lines of Fadeev $^{8}$,
Sukumar $^{9}$, or Kosteleck\'y and Nieto $^{10}$, i.e., as changing the
angular momentum, is a delicate point, and actually an important open issue in
Physics. Indeed, to associate the DW procedure with a change $n\rightarrow n-1$
at fixed $l$ as suggested by Haymaker and Rau seems more natural. Therefore,
I shall tackle this point as well.

The idea is to pass from the half line case to the full line one by the Langer
change
of variable $x=\ln \rho$ and wave function  $\phi (x)=\exp (-x/2)u$.
Then the standard DO Schr\"odinger equation at zero energy turns into the
Rosen-Morse (RM) problem
$$
\Bigg[-\frac{d^2}{dx^2}+(n-\frac{1}{2})^2-\frac{(n-\frac{1}{2})(n+\frac{1}{2})}
{\cosh ^2x}\Bigg ]\phi =0
\eqno(17)
$$
where I made use of $l+1=n$, considered as the lowest eigenvalue and fixing
the zero of the energy scale on it. I shall also make the shift
$n_b=n+1/2$, taking only the integer part $[n_b]$, since I am not
interested in the WKB analysis and I need the standard form of the potential
$$
V^{-}=-\frac{n_b(n_b+1)}{\cosh ^2x}~.
\eqno(18)
$$
Then, Eq.~(17) is just the RM eigenvalue problem with
$[n_b]$ bound states, of eigenvalues $-[n_b]^2$.
The particular superpotential is $W=n_b\tanh x$ and the superpartner is
$V^{+}=-\frac{n_b(n_b-1)}{\cosh ^2x}$
and therefore has one bound state less than $V^{-}$.
The supersymmetry of the full line RM potential is well established,
including
the strictly isospectral issue, see, e.g., $^{11}$. Multiple-parameter
families of strictly isospectral $-\cosh ^{-2}x$ potentials can be obtained
easily and a nice picture of the translational motion of the wells, each with
one bound state, towards $\pm \infty$ as
the family parameters are going to the Abraham-Moses limit ($-1^{-}$)
and Pursey limit ($0^{+}$), respectively, is also well known.
For example, in the case of a single bound state, the one-parameter family
of $-\cosh ^{-2}x$ potentials reads
$$
V_0(\lambda _0)=-2\cosh ^{-2}(x+\frac{1}{2}\ln(1+1/\lambda _0))
\eqno(19)
$$
where one can see the parameter imbedded in the argument of a logarithm,
adding to the argument of the hyperbolic function, which is a general
feature of this case and the reason for the translation of the potentials.

Exactly as suggested by Haymaker and Rau, after all the supersymmetric job
on the full line, one should go back to the half line to obtain the right
superpartner and isospectral families.
By doing so, one gets the DO particular superpotential
$W=(\frac{1}{2}-n)\xi(\rho)$, a superpartner effective potential
$U^{+}=l(l+1)/\rho ^2-
(2l+1)(2l-1)/(1+\rho ^2)^2$, whereas the effect of the isospectral
parameter shows up in a rescaling of the $\rho$ variable of the type
$\rho _{\lambda _0}=\sqrt{1+\frac{1}{\lambda _0}}\rho _{\infty}$, where
$\rho _{\infty}$ is the initial coordinate, corresponding to
$\lambda _0=\infty$. In other words, we have a
rescaling of the radius of the fish-eye lens occuring simultaneously with
the change $n\rightarrow n-1$. The fish-eye lens of radius $R_{\lambda _0}=
R\sqrt{\lambda _0/(\lambda _0 +1)}$ has a zero-bound state of degeneracy
reduced by one unit with respect to the lens of radius $R$.
This is a counterpart of what happens in the hydrogen case,
where one has the simultaneous change $n\rightarrow n-1$ and $Z\rightarrow
Z(1-1/n)$.$^{7}$

In the atomic aufbau case the same results hold with only minor changes in
the RM problem, which this time turns to apply to the potential
$-N(N+1)/4\cosh ^2(x/2)$ of eigenvalues $-N^2/4$, where $N=2l+1$.

\section{{\bf CONCLUSION}} 

In conclusion, I presented a rather detailed discussion of the DO
wave problem in the nodeless radial sector by employing DW tools,
with special focus on the fish-eye lens case.
While the $l$-changing half line procedure does not look very natural,
the $N$-changing full line one appears to be
more acceptable. Of course, it is more convenient not to face the centrifugal
singularity by sending it to $-\infty$, avoiding all related difficulties.
One just pass in this way to an easy to handle reflectionless and
shape invariant RM problem.
The centrifugal singularity is not ``active".
Moreover, the isospectral parameter plays
different roles, as a strong damping in the half line case, and as a
``dilatation'' parameter when the full line procedure is used.

As an important by-product of this study I introduced generalized l- and
$\lambda$-dependent families of Maxwell indices of refraction, which
presumably
might be related to surface-scattering optical phenomena. Indeed, the squares
of the radial factors are peaked close to the surface $\rho=1$.
Finally, as one might guess from the above analysis,
there might be potential applications of DW techniques in graded index optics.


\nonumsection{Acknowledgements}
\noindent
This work was partially supported by the Consejo Nacional de Ciencia y
Tecnolog\'{\i}a (M\'exico) (CONACyT) Project No. 4868-E9406.
I thank Kurt Bernardo Wolf for asking me if it is not possible to
have more than one type of Maxwell fish-eye lenses.

\bigskip

\nonumsection{Figure Captions}

{\bf Fig.~1.} The original Maxwell refractive index $n_M$,
the ``isospectral" one $n_{iso}$, their relative ratio $\frac{n_{iso}}{n_M}-1$,
and the square of the radial factor $f$ [Eq.~(16)] for
(a) $l=1$, $\lambda=1$ and (b) $l=1$, $\lambda=10$, respectively.

{\bf Fig.~2.} The same functions for (a) $l=2$, $\lambda=1$ and (b) $l=2$,
  $\lambda=10$, respectively.

In both figures, I have used $n_{iso}/n_M=1+\frac{1}{2}
\frac{V_{\lambda}}{V_{M}}$, where $V_{\lambda}$ is the negative of the
$\lambda$-depending part in Eq.~(14), whereas $V_{M}$ is the negative of the
first term therein. Since $V_{\lambda}$ may be negative,
the relative ratio may be negative as well.
The indices have been normalized to $(l+1/2)^2$ to
be as close as possible to the Maxwell normalization constant $n(0)=2$, and 
also to include the case $l=0$.

\newpage

\nonumsection{References}



\begin{thebibliography}{000}


\bibitem{1}
Yu.N. Demkov and V.N. Ostrovsky, Zh. Eksp. Teor. Fiz.
{\bf 60}, 2011 (1971)
[JETP {\bf 33}, 1083 (1971)].

\bibitem{2}
H.C. Rosu, M. Reyes, K.B. Wolf, and O. Obreg\'on,
Phys. Lett. A {\bf 208}, 33
(1995); in {\it Second Iberoamerican Meeting on Optics}, edited by
D. Malacara, S.E. Acosta, R. Rodr\'{\i}guez Vera, Z. Malacara, and A. Morales,
SPIE Proc. Vol. 2730, (SPIE, Bellingham, WA, 1996), pp 436-439.

\bibitem{3}
Y. Kitagawara and A.O. Barut, J. Phys. B {\bf 16}, 3305 (1983);
{\bf 17}, 4251 (1984). [See also, Yu.N. Demkov and V.N. Ostrovsky,
Zh. Eksp. Teor. Fiz. {\bf 62}, 125 (1971); Errata {\bf 63}, 2376 (1972)
[Sov. Phys. JETP {\bf 35}, 66 (1972)]; V.N. Ostrovsky, J. Phys. B {\bf 14},
4425 (1981); V.N. Ostrovsky, Latin-American School of Physics XXX ELAF,
eds. O. Casta\~nos, R. L\'opez-Pe\~na, J.G. Hirsch, K.B. Wolf, AIP Conf. Proc.
{\bf 365}, pp 191-216 (1996)].

\bibitem{4}
E. Witten, Nucl. Phys. B {\bf 185}, 513 (1981).


\bibitem{5}
I.S. Gradshteyn and I.M. Ryzhik, {\em Table of integrals, series,
and products},
4th ed. (Academic, New York, 1980), pp. 136, 134, 131, and 130.

\bibitem{6}
A. Khare and U.P. Sukhatme, Phys. Rev. A {\bf 40}, 6185 (1989).

\bibitem{7}
R.W. Haymaker and A.R.P. Rau, Am. J. Phys. {\bf 54}, 928 (1986).

\bibitem{8}
L.D. Fadeev, J. Math. Phys. {\bf 4}, 72 (1963).

\bibitem{9}
C.V. Sukumar, J. Phys. A {\bf 18}, 2917 (1985); {\bf 18}, 2937 (1985).

\bibitem{10}
V.A. Kosteleck\'y and M.M. Nieto, Phys. Rev. Lett. {\bf 53}, 2285 (1984).

\bibitem{11}
W.-Y. Keung, U.P. Sukhatme, Q. Wang, and T.D. Imbo, J. Phys. A
{\bf 22}, L987 (1989).












\end{thebibliography}
\end{document}